\newcommand*{\wn}{cm$^{-1}$}
\newcommand*{\AX}{A$^1\Pi-$X$^1\Sigma^+$}
\newcommand*{\dX}{d$^3\!\Delta-$X$^1\Sigma^+$}
\newcommand*{\eX}{e$^3\Sigma^--$X$^1\Sigma^+$}

\newcommand*{\BA}{B$^1\Sigma^+-$A$^1\Pi$}
\newcommand*{\BX}{B$^1\Sigma^+-$X$^1\Sigma^+$}
\newcommand*{\Be}{B$^1\Sigma^+-$e$^3\Sigma^-$}
\newcommand*{\Bd}{B$^1\Sigma^+-$d$^3\!\Delta$}

\newcommand*{\ds}{d$^3\!\Delta$}
\newcommand*{\es}{e$^3\Sigma^-$}
\newcommand*{\as}{a$'^3\Sigma^+$}
\newcommand*{\Ds}{D$^1\!\Delta$}
\newcommand*{\Is}{I$^1\Sigma^-$}
\newcommand*{\As}{A$^1\Pi$}
\newcommand*{\Bs}{B$^1\Sigma^+$}
\newcommand*{\Xs}{X$^1\Sigma^+$}
\newcommand*{\0}{$v=0$}

\newcommand*{\CO}{$^{13}$C$^{16}$O}

\documentclass{tMPH2e}
\usepackage{graphics}
\usepackage{dcolumn}
\usepackage[version=3]{mhchem}
\usepackage[english]{babel}
\usepackage{dcolumn}
\usepackage{graphicx}
\usepackage{amsmath}
\usepackage{tabularx}
\usepackage{bigstrut}
\usepackage{longtable}
\usepackage{rotating}
\usepackage{comment}
\usepackage{float}
\usepackage{verbatim}
\usepackage{color, soul}
\usepackage{threeparttable}%The solution to the footnote problem in the tabular environment. Can be used inside a table environment. Make sure the caption is placed inside the table environment and not the threeparttable environment, otherwise the caption is made the same width as the table and the \centering command does not work.
\usepackage{threeparttablex}%The solution to the footnote problem in the longtable environment. This seemed to work well, although the environment is slightly different to the normal threeparttable environment! Also had some problems with the table when it is not split over a page! It seems to want to put a header and a footer on the previous page?! The only solution I could find was to place the table in the text where it is split for sure!

\begin{document}

\title{Spectroscopy and perturbation analysis of the $\bm{{\rm A}{}^1\Pi(v=0)}$ state of $\bm{{}^{13}{\rm C}{}^{16}{\rm O}}$}

\author{
M.~L.~Niu,$^{a}$
R.~Hakalla,$^{b}$
T.~Madhu~Trivikram,$^{a}$
A.~N.~Heays,$^{a}$
N.~de~Oliveira,$^{c}$
E.~J.~Salumbides,$^{a}$
and W.~Ubachs$^{a}$
\\\vspace{6pt}%
$^{a}${\em{Department of Physics and Astronomy, and LaserLaB, Vrije Universiteit, De Boelelaan 1081, 1081 HV Amsterdam, The Netherlands}}\\%
$^{b}${\em{Materials Spectroscopy Laboratory, Department of Experimental Physics, Faculty of Mathematics and Natural Science, University of Rzesz\'{o}w, ul. Prof. S. Pigonia 1, 35-959 Rzesz\'{o}w, Poland.}}\\%
$^{c}${\em{Synchrotron SOLEIL, Orme de Merisiers, St. Aubin, BP 48, F-91192 Gif sur Yvette Cedex, France.}}%
\\\vspace{6pt}%
}

\date{\today}

\maketitle

\begin{abstract}
  \label{abstract}
The lowest $v=0$ level of the \As\,state of the $^{13}$C$^{16}$O isotopologue of carbon monoxide has been reinvestigated with a variety of high resolution spectroscopic techniques. The \AX$(0,0)$ band has been studied by vacuum-ultraviolet Fourier-transform absorption spectroscopy, using the SOLEIL synchrotron as a radiation source. Spectra were obtained under quasi-static gas conditions at liquid-nitrogen temperature, room temperature and at an elevated temperature of 900 K, with absolute accuracies of 0.01$-$0.03 \wn. Two-photon Doppler-free laser spectroscopy has been applied to a limited number of transitions in the \AX$(0,0)$ band, under collision-free circumstances of a molecular beam, yielding an absolute accuracy of 0.002 \wn. The third technique is high-resolution Fourier-transform emission spectroscopy in the visible region applied to the \BA$(0,0)$ band in a gas discharge, at an absolute accuracy of up to 0.003 \wn. With these methods rotational levels of  \As\,(0) could be studied in both parity components up to a rotational quantum number of $J=46$.
  The frequencies of 397 transitions were used to analyse the perturbations between the \As$(0)$ level by vibrational levels of the \Ds, \es, \ds, and \as\ states.
  % \bigskip
\end{abstract}


\section{Introduction}

Spectroscopy of the carbon monoxide molecule remains of interest, in large part, because of its role as the prime tracer of astrophysical molecular gas in interstellar clouds~\cite{Kong2015}, circumstellar material~\cite{Flaherty2015}, entire galaxies~\cite{Bolatto2013}, at very high redshifts~\cite{Levshakov2012} and in gamma-ray bursts~\cite{Prochaska2009}.
Astronomical CO has also been observed in the ultraviolet, through absorption or emission of the fourth positive system (\AX) \cite{Federman2003,Mcjunkin2013}.
In addition, the CO molecule is of interest from a molecular physics perspective, since its spectrum exhibits a wealth of perturbations arising from spin-orbit and centrifugal effects, as well as Rydberg-valence mixing \cite{krupenie1966,Tilford1972,eidelsberg2004a}.

Perturbations affecting the \As{} state were extensively investigated by Simmons and Tilford~\cite{Simmons1966} and analyzed in the benchmark study of Field \emph{et al.}~\cite{Field-thesis,Field1972a} and studied many times thereafter in its dominant isotopologue $^{12}$C$^{16}$O and minor variants (e.g., \cite{Haridass1994,Eikema1994,Jolly1997,Kepa2011}).
Recently the spectrum of \AX\ in  $^{12}$C$^{16}$O was reinvestigated by Doppler-free laser methods~\cite{Niu2015} and vacuum-ultraviolet Fourier-transform (VUV-FT) spectroscopy~\cite{Niu2013,Niu2016}, both at high resolution.
The \AX{} system of the  $^{13}$C$^{16}$O isotopologue was investigated in detail by Haridass and Huber~\cite{Haridass1994}, after the observation of unresolved band heads by Tilford and Simmons \cite{Tilford1972}. Recently, a high-resolution study of the \AX\ ($v'$,0)  bands for $v'=0-9$ was reported, based on measurements with the VUV-FT instrument at the SOLEIL synchrotron~\cite{Gavilan2013}.
Further observations of \As$(v)$ levels for this isotopologue were made in visible emission via the \BA\ \AA ngstr\"{o}m bands by the Rzesz\'{o}w group~\cite{Kepa2003,Kepa2011,Hakalla2012}, while the \BX\ system was investigated by two groups with improved accuracy~\cite{Tilford1968,Eidelsberg1987}. Higher lying states for the $^{13}$C$^{16}$O isotopologue have also been investigated~\cite{Eidelsberg1992,Cacciani1995}, while the level structure of the \Xs\ ground state is known to great precision~\cite{George1994}.

The present study aims to re-investigate the spectroscopic structure of the \0\ vibrational level of the \As{} state for the $^{13}$C$^{16}$O isotopologue with the highest-possible accuracy. For this purpose, high resolution Doppler-free laser measurements of the \AX$(0,0)$ band are combined with VUV-FT measurements under various temperature conditions that permit access to higher rotational levels, as well as with high-resolution visible Fourier-transform (VIS-FT) emission spectra obtained from a discharge. These data are combined in a comprehensive deperturbation analysis with the purpose of quantitatively describing the \As\,(0) level and various vibronic states with which it interacts.

\section{Experiments}

Independent high-resolution studies of the \AX$(0,0)$ band of $^{13}$C$^{16}$O were performed by laser and VUV-FT spectroscopy, while the \BA$(0,0)$ emission band was investigated with VIS-FT spectroscopy from a discharge. The methods used and the results obtained in the three different experiments are described below.

\subsection{Laser experiments}
\label{sec:laser experiments}

\begin{figure}
  \begin{center}
    \includegraphics[width=\linewidth]{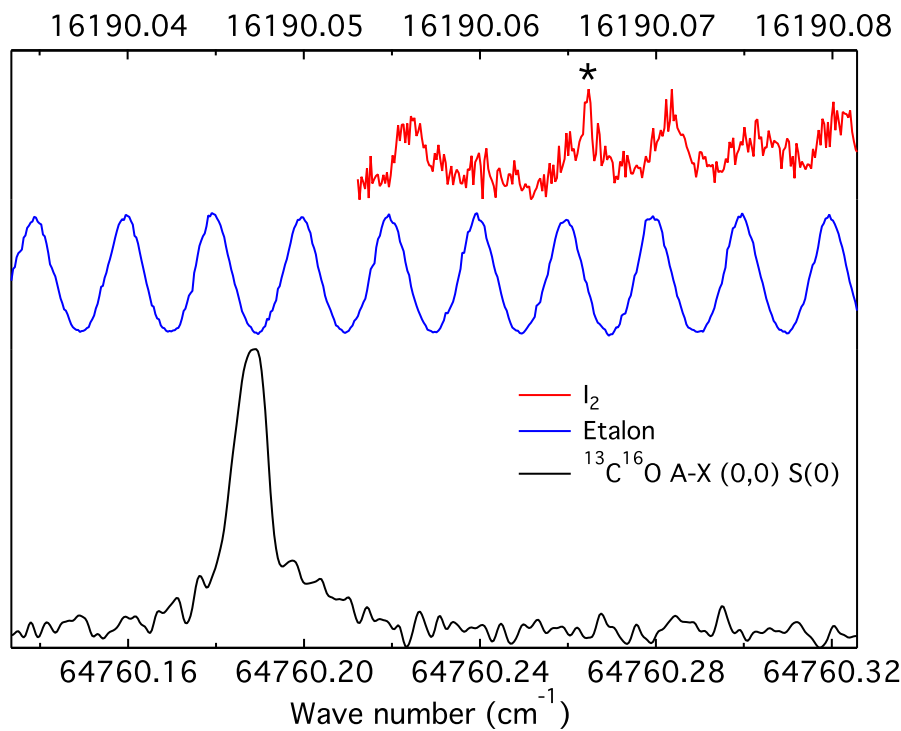}
    \caption{ The S(0) transition in the \AX$(0,0)$ band of $^{13}$C$^{16}$O, as measured via two-photon Doppler-free laser spectroscopy, plotted as a black line. The red and blue lines show the saturated iodine spectrum and etalon markers used for calibration and interpolation. The asterisk (*) indicates the hyperfine component of the iodine line used for the absolute wave number calibration, in this case with overlapping contributions from the $a_5$ and $a_6$ components of the B$-$X$(8,2)$ $R(81)$ line at $16\,190.06605$ and $16\,190.06616$ \wn~\cite{Bodermann2002,Xu2000}. The upper scale represents the fundamental wave number of the cw-seed laser, and the lower scale the two-photon excitation energy.}
    \label{fig:PDAspec}
  \end{center}
\end{figure}

Figure~\ref{fig:PDAspec} shows a typical spectrum of the S(0) transition of \AX$(0,0)$ band of $^{13}$C$^{16}$O obtained in the laser experiment.
Here, a narrow-band pulsed dye amplifier (PDA) laser system is used~\cite{Ubachs1997}, seeded by the output of a cw ring dye laser which is optimized for operation at 618 nm, and pumped by a Nd:YAG laser with a repetition rate of 10 Hz.
The pulsed output of the PDA is frequency doubled and then employed in a two photon experiment with counter-propagating laser beams aligned in the geometry of a Sagnac interferometer~\cite{Hannemann2007} to avoid a residual Doppler shift. The laser beams crossed the 99\% $^{13}$C-enriched CO molecular beam, perpendicularly.
A second pulsed laser at 207 nm, based on a frequency-tripled dye laser, delayed by around 10 ns from the probe laser to minimize the AC-Stark effects, is employed for ionizing the \As\ excited state population. This method of $2+1'$ resonant-enhanced multiphoton ionization (REMPI) yielded better results, than one-color $2+2$ REMPI measurements, in terms of signal-to-noise ratio and suppression of the AC-Stark effect.

\begin{figure}
  \begin{center}
    \includegraphics[width=\linewidth]{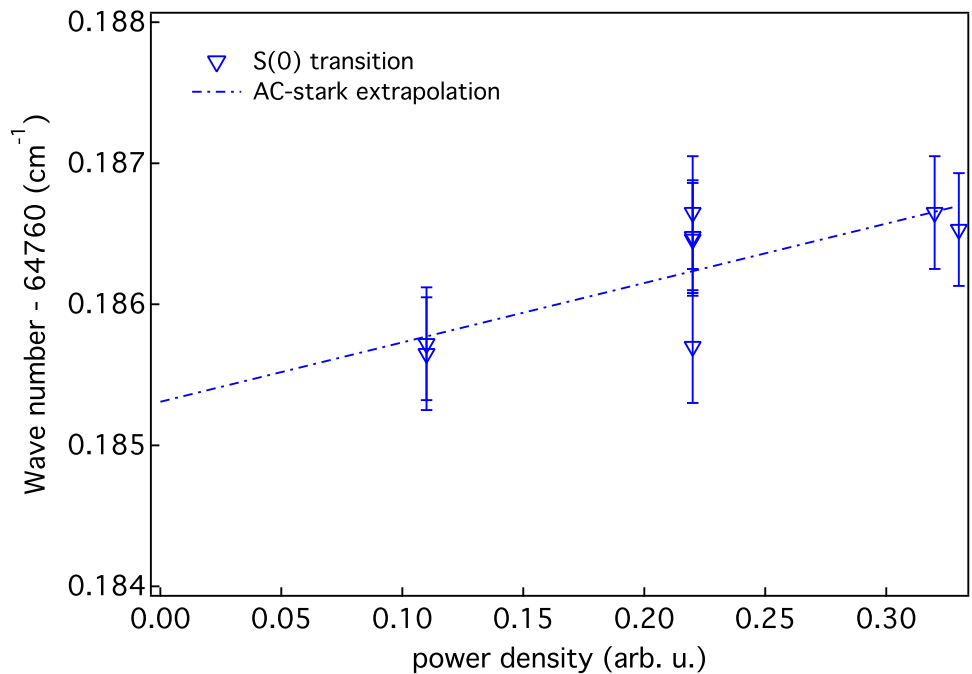}
    \caption{Wave number of the S(0) transition of the \AX$(0,0)$ band of $^{13}$C$^{16}$O extrapolated to zero probe-laser power, yielding an AC-Stark shift corrected value.}
    \label{fig:stark}
  \end{center}
\end{figure}

In order to reduce the AC-Stark effect induced by the probe laser, the resonant CO lines were measured for different laser intensities and their transition frequencies extrapolated to a zero-intensity AC-Stark-free frequency. An example of such an extrapolation, for the S(0) line, is included in Fig.~\ref{fig:stark}.
Using part of the cw-seed radiation, an etalon spectrum and a saturated iodine spectrum are recorded simultaneously with the $^{13}$C$^{16}$O two-photon spectrum to interpolate and linearize the scan and to calibrate the absolute frequency of the resonances, respectively.
The frequency chirp between the PDA output and cw seed laser is measured on-line, yielding an average chirp offset that is included in the frequency calibration.
% All laser experiments were performed using 99\% enriched samples of $^{13}$CO gas.
For further details on the two-photon Doppler-free laser experiments on CO we refer to previous publications~\cite{Niu2013,Niu2015}.

The sources of transition-frequency uncertainty in the laser experiment were investigated. The statistical uncertainty was estimated from multiple measurements of the observed lines to be 30~MHz.
% A dominant source of systematic uncertainty is due to residual frequency chirp, even after correcting for this phenomenon quantitatively, and is 8~MHz.
The frequency chirp of the fundamental radiation was measured to be $-11\pm 2$\,MHz, leading to a $-44$\,MHz correction to the measured \AX{} frequencies with a residual uncertainty of 8\,MHz.
The uncertainty of the AC-Stark-shift extrapolation is 30~MHz. Other sources of uncertainty, such as line-fitting, I$_2$ calibration, etalon calibration, DC-Stark shift and residual Doppler shift, make much smaller contributions and can be ignored. The overall uncertainty ($1\sigma$) of the two-photon resonant frequencies of $^{13}$C$^{16}$O is then conservatively estimated to be 0.002~\wn.

The transition frequencies of six \CO\ lines were measured, calibrated, and corrected for chirp and AC-Stark effects, with final values listed in Table~\ref{AXlasertransition}. 

\begin{table}
  \begin{center}
    \caption{Results from the two-photon Doppler-free laser experiment on the \AX\ $(0,0)$ band of $^{13}$C$^{16}$O. Transition frequencies (in vacuum \wn). The $1\sigma$ uncertainty of all frequencies is better than 0.002\,cm$^{-1}$.}
    \label{AXlasertransition}
    \begin{tabular}
      {l@{\hspace{50pt}}r}
      \hline
      \hline\\[-2ex]
      Line	&	Frequency	\\
      \hline\\[-2ex]
      S(0)	&	64760.185	\\
      S(1)	&	64765.528	\\
      R(1)	&	64756.560	\\
      Q(1)	&	64750.481	\\
      Q(2)	&	64749.157	\\
      P(2)	&	64743.144	\\
      \hline
      \hline
    \end{tabular}
  \end{center}
\end{table}

\subsection{VUV-FT synchrotron experiments}
The \AX\ $(0,0)$ single-photon absorption band of $^{13}$C$^{16}$O was measured using the VUV Fourier-transform spectrometer at the SOLEIL synchrotron.
The operation of this device and its use for molecular spectroscopy have been well described~\cite{Deoliveira2011,Niu2016}.
All measurements were performed under steady-state gas conditions, whereby gas is continuously introduced to the interaction region and flows out of a windowless cell. Recordings were made with the cell cooled to nearly liquid-nitrogen temperature (yielding an effective gas temperature of 90\,K), at room temperature (295\,K), and in a specially-designed high-temperature cell (900\,K)~\cite{Niu2015a}. The VUV-FT measurements were performed using 99\% enriched samples of $^{13}$C$^{16}$O gas. The sampling rate and maximum path difference of the FT spectrometer was set to provide resolutions of 0.075, 0.075, and 0.27\,\wn\ FWHM for the 90, 295, and 900\,K measurements, respectively.
The respective Doppler broadening of these measurements are 0.08, 0.14, and 0.25\,\wn\,FWHM. 
Further experimental and analytical details of the VUV-FT spectra are provided in Refs.~\cite{Niu2013,Niu2015,Niu2016}.

\begin{figure}
  \begin{center}
    \includegraphics[width=\linewidth]{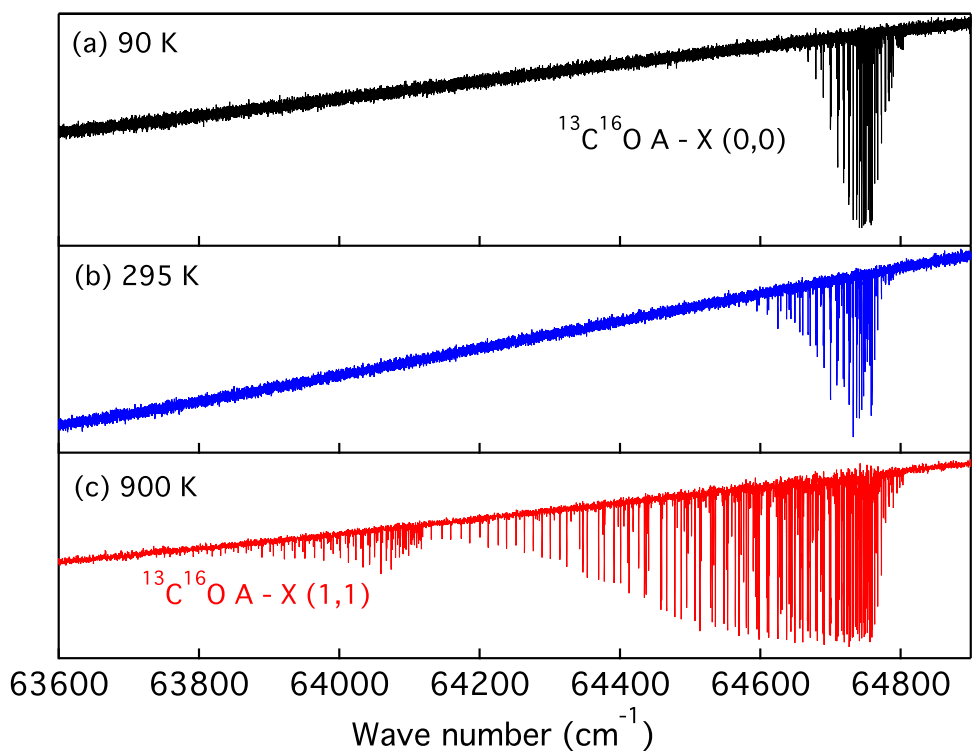}
     \caption{Spectra of the \AX$(0,0)$ band of $^{13}$C$^{16}$O recorded with the VUV-FT at the SOLEIL synchrotron under three different temperature regimes: (a) cooling of the cell with liquid nitrogen; (b) at room temperature; (c) cell heated to $\sim 900$ K. The A$^1\Pi-$X$^1\Sigma^+(1,1)$ absorption band is also evident at 64100\,\wn{} in the latter case.}
    \label{fig:FTAXspec}
  \end{center}
\end{figure}

\begin{figure*}[htbp]
  \begin{center}
    \includegraphics[width=\linewidth]{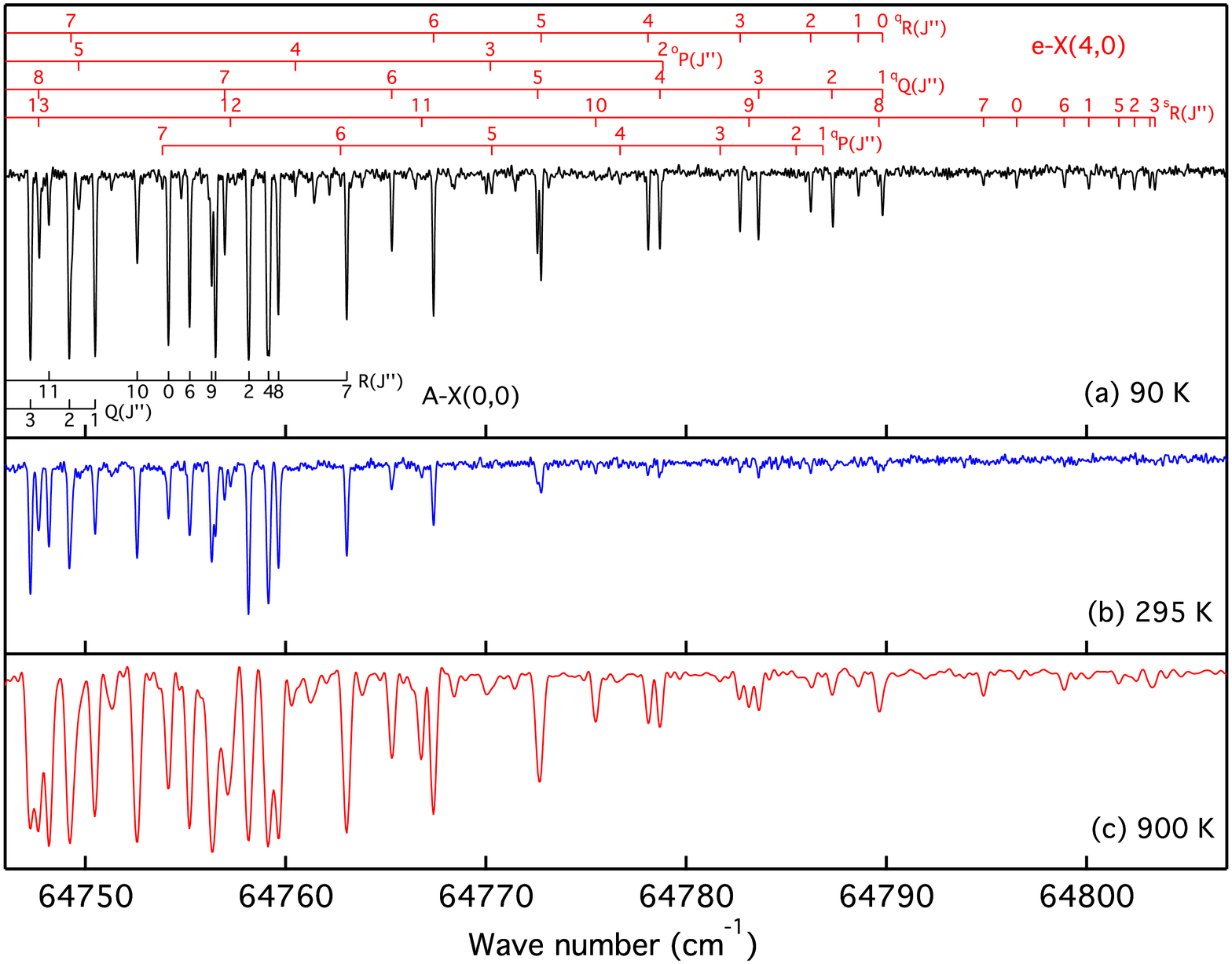}
    \caption{Detail spectra of the \AX$(0,0)$ bandhead region and transitions to the perturbing \eX$(4,0)$ state. Measured with the VUV-FT spectrometer under three temperature regimes: (a) cell cooled with liquid nitrogen to 90\,K; (b) cell at room temperature; (c) cell heated to $\sim 900$ K. Unassigned lines are due to $^{13}$C$^{18}$O contamination.}
    \label{fig:FTAXspecdetail}
  \end{center}
\end{figure*}

Overview spectra of the \AX$(0,0)$ band, recorded at the three different temperatures, are displayed in Fig.~\ref{fig:FTAXspec}, covering $64\,000-64\,800$ \wn.
These spectra enable good line-position determinations for transitions from different ranges of ground state $J''$. The lowest $J''$ states are populated at 90\,K and lead to transitions with the narrowest Doppler widths, resulting in the least-congested spectra (i.e., in Fig.~\ref{fig:FTAXspec}(a)), thus facilitating straightforward line assignment. Analysing the 90\,K spectrum first aided the assignment of higher temperature spectra in Fig.~\ref{fig:FTAXspec}(b) and (c). Intermediate $J''$ levels have higher populations at 295\,K and also provide a cross-check for low $J''$ lines.
The 900\,K spectrum presented in Fig.~\ref{fig:FTAXspec}(c) was measured at about 60 times higher column density than the room temperature spectrum, with the goal of obtaining as much information as possible on the highest rotational quantum states, in which case transitions from the low $J''$ were saturated. Ultimately, the rotational progression of \AX$(0,0)$ could be followed to $64\,000$ \wn\ and for transition with $J'$ as high as 46, whereupon a severe overlap with the \AX$(1,1)$ band sets in.

Figure \ref{fig:FTAXspecdetail} shows details of VUV-FT spectra in the bandhead region and demonstrates the complementarity of spectra recorded under different temperature and saturation conditions.
Also appearing are forbidden transitions to the \es$(4)$ level due to intensity borrowing from the \AX$(0,0)$ band. The mutual perturbation of \es$(4)$ and \As$(0)$ also leads to rotational line structure discontinuities as, for instance, the perturbation of $R(6)$ and $R(7)$ lines in Fig.~\ref{fig:FTAXspecdetail}.
Further lines appear from the forbidden \dX$(4,0)$ band in our spectra.

The frequency calibration of Fourier-transform spectra is inherently linear and requires only a single absolute standard.
Here, the absolute calibration was made by reference to the results presented in Table~\ref{FTandlasertransition} while accounting for the different \As{} $\Lambda$-doubling components accessed by one- and two-photon transitions, leading to an absolute calibration uncertainty of VUV-FT of 0.01\,\wn.
The frequencies of strongly absorbed, unsaturated, and unblended \AX{} transitions were determined with a $1\sigma$ random uncertainty of 0.01~\wn\,and with somewhat greater uncertainties for weak and blended lines and those measured at 900\,K temperature, leading to total frequency uncertainties that vary between 0.01 and 0.03~\wn.

Transition vacuum wave numbers resulting from the combined analysis of all three \AX$(0,0)$ spectra are presented in Table~\ref{FTandlasertransition} and vacuum wave numbers of the perturber lines are listed in Table~\ref{FT-pert}.

\begin{table*}
  \begin{center}
    \caption{Transition frequencies (in vacuum \wn) of the $^{13}$C$^{16}$O \AX\ $(0,0)$ band obtained in the VUV-FT experiment and \BA\ $(0,0)$ band obtained in the VIS-FT experiment. $J''$ is the rotational quantum number of the lower state and subscripts $e$ and $f$ indicate the \As\ state parity. The superscripts \textit{b} and \textit{w} indicate blended and weak transitions, respectively.
      Frequencies are given to three decimal places when known to better than 0.01\,\wn{} and to two places when less accurate.
    }
    \label{FTandlasertransition}
    \renewcommand\arraystretch{1.1}
    \small
    \begin{tabular}
      {c@{\hspace{15pt}}l@{\hspace{7pt}}l@{\hspace{7pt}}l@{\hspace{15pt}}l@{\hspace{7pt}}l@{\hspace{7pt}}l}
      \hline
      \hline\\[-2ex]
   &\multicolumn{3}{c}{\AX$(0,0)$}	&	\multicolumn{3}{c}{\BA$(0,0)$}\\
      \cline{2-7}\\[-2ex]

$J'' $ & R$_e(J'')  $     & Q$_f(J'')  $     & P $_e(J'')   $   & R $_e(J'')   $    & Q $_f(J'')   $    & P$_e(J'')$       \\
\hline\\[-2ex]
0      & 64754.15$^{ }$   &                  &                  &                   &                   &                  \\                        
1      & 64756.51$^{ }$   & 64750.49$^{ }$   &                  & 22173.83$^{ wb }$ & 22166.35$^{ wb }$ & 22162.65$^{ wb }$\\     
2      & 64758.16$^{ b }$ & 64749.21$^{ }$   & 64743.12$^{ }$   & 22178.97$^{ w }$  & 22167.744$^{ }$   & 22160.34$^{ wb }$\\     
3      & 64759.15$^{ b }$ & 64747.26$^{ }$   & 64738.13$^{ }$   & 22184.84$^{ b }$  & 22169.833$^{ }$   & 22158.779$^{ }$  \\       
4      & 64759.15$^{ b }$ & 64744.66$^{ }$   & 64732.48$^{ b }$ & 22191.52$^{ b }$  & 22172.634$^{ }$   & 22158.00$^{ b }$ \\      
5      & 64758.17$^{ b }$ & 64741.36$^{ }$   & 64726.02$^{ }$   & 22199.07$^{ b }$  & 22176.17$^{ b }$  & 22158.105$^{ }$  \\       
6      & 64755.22$^{ }$   & 64737.34$^{ }$   & 64718.76$^{ }$   & 22207.806$^{ }$   & 22180.496$^{ }$   & 22159.396$^{ }$  \\       
7      & 64763.06$^{ }$   & 64732.52$^{ b }$ & 64710.37$^{ }$   & 22218.46$^{ b }$  & 22185.67$^{ b }$  & 22162.61$^{ b }$ \\      
8      & 64759.65$^{ }$   & 64726.71$^{ }$   & 64700.15$^{ b }$ & 22218.42$^{ b }$  & 22191.857$^{ }$   & 22155.115$^{ }$  \\       
9      & 64756.31$^{ }$   & 64719.65$^{ }$   & 64700.60$^{ }$   & 22229.650$^{ }$   & 22199.358$^{ }$   & 22158.920$^{ }$  \\       
10     & 64752.59$^{ }$   & 64727.84$^{ }$   & 64689.84$^{ }$   & 22240.855$^{ }$   & 22191.658$^{ }$   & 22162.69$^{ b }$ \\      
11     & 64748.19$^{ }$   & 64718.28$^{ }$   & 64679.17$^{ }$   & 22252.500$^{ }$   & 22201.749$^{ }$   & 22166.90$^{ b }$ \\      
12     & 64742.78$^{ }$   & 64709.17$^{ }$   & 64668.11$^{ }$   & 22264.87$^{ b }$  & 22211.44$^{ b }$  & 22171.839$^{ }$  \\       
13     & 64735.65$^{ }$   & 64700.15$^{ b }$ & 64656.38$^{ }$   & 22278.276$^{ }$   & 22221.119$^{ }$   & 22177.82$^{ b }$ \\      
14     & 64738.93$^{ }$   & 64690.94$^{ }$   & 64643.64$^{ }$   & 22293.476$^{ }$   & 22230.99$^{ b }$  & 22185.60$^{ b }$ \\      
15     & 64731.21$^{ }$   & 64681.32$^{ }$   & 64629.19$^{ }$   & 22298.291$^{ }$   & 22241.339$^{ }$   & 22182.990$^{ }$  \\       
16     & 64723.66$^{ }$   & 64671.21$^{ }$   & 64625.13$^{ b }$ & 22314.14$^{ b }$  & 22252.222$^{ }$   & 22191.44$^{ b }$ \\      
17     & 64715.83$^{ }$   & 64660.56$^{ }$   & 64610.10$^{ }$   & 22329.876$^{ }$   & 22263.690$^{ }$   & 22199.76$^{ b }$ \\      
18     & 64707.52$^{ }$   & 64649.35$^{ }$   & 64595.24$^{ }$   & 22345.94$^{ b }$  & 22275.778$^{ }$   & 22208.424$^{ }$  \\        
19     & 64698.68$^{ }$   & 64637.52$^{ }$   & 64580.12$^{ }$   & 22362.514$^{ }$   & 22288.511$^{ }$   & 22217.605$^{ }$  \\       
20     & 64689.16$^{ }$   & 64625.13$^{ b }$ & 64564.48$^{ }$   & 22379.691$^{ }$   & 22301.90$^{ b }$  & 22227.389$^{ }$  \\       
21     & 64678.96$^{ }$   & 64612.00$^{ }$   & 64548.32$^{ }$   & 22397.515$^{ }$   & 22315.988$^{ }$   & 22237.828$^{ }$  \\       
22     & 64667.27$^{ }$   & 64598.22$^{ }$   & 64531.55$^{ }$   & 22416.068$^{ }$   & 22330.844$^{ }$   & 22249.033$^{ }$  \\       
23     & 64657.68$^{ }$   & 64582.92$^{ }$   & 64514.02$^{ b }$ & 22436.211$^{ }$   & 22347.208$^{ }$   & 22261.788$^{ }$  \\        
24     & 64645.09$^{ }$   & 64569.68$^{ }$   & 64495.08$^{ }$   & 22454.294$^{ }$   & 22361.614$^{ }$   & 22272.455$^{ }$  \\       
25     & 64631.27$^{ }$   & 64553.48$^{ }$   & 64478.24$^{ }$   & 22475.39$^{ wb }$ & 22379.034$^{ }$   & 22286.216$^{ }$  \\       
26     & 64628.02$^{ }$   & 64536.04$^{ }$   & 64458.36$^{ }$   & 22497.74$^{ wb }$ & 22397.713$^{ }$   & 22301.24$^{ wb }$\\     
27     & 64609.48$^{ }$   & 64529.18$^{ }$   & 64437.28$^{ }$   & 22509.57$^{ w }$  & 22405.86$^{ w }$  & 22305.70$^{ wb }$\\     
28     & 64593.44$^{ }$   & 64507.00$^{ }$   & 64426.77$^{ }$   & 22536.73$^{ w }$  & 22429.36$^{ wb }$ & 22325.54$^{ w }$ \\      
29     & 64576.38$^{ }$   & 64487.34$^{ }$   & 64400.98$^{ }$   &                   & 22450.38$^{ wb }$ & 22342.90$^{ wb }$\\     
30     & 64572.75$^{ }$   & 64466.67$^{ }$   & 64377.71$^{ }$   &                   & 22472.45$^{ wb }$ & 22361.34$^{ w }$ \\      
31     & 64551.80$^{ }$   & 64459.40$^{ }$   & 64353.41$^{ }$   &                   & 22481.16$^{ wb }$ &                  \\                        
32     & 64533.62$^{ }$   & 64434.84$^{ }$   & 64342.57$^{ }$   &                   &                   &                  \\                        
33     & 64515.65$^{ }$   & 64413.03$^{ }$   & 64314.39$^{ }$   &                   &                   &                  \\                        
34     & 64497.32$^{ }$   & 64391.62$^{ }$   & 64289.02$^{ }$   &                   &                   &                  \\                        
35     & 64478.38$^{ }$   & 64369.64$^{ }$   & 64263.85$^{ }$   &                   &                   &                  \\                        
36     & 64459.40$^{ b }$ & 64347.23$^{ }$   & 64238.34$^{ }$   &                   &                   &                  \\                        
37     & 64439.34$^{ }$   & 64324.20$^{ b }$ & 64212.20$^{ }$   &                   &                   &                  \\                        
38     & 64418.75$^{ }$   & 64300.72$^{ }$   & 64185.99$^{ }$   &                   &                   &                  \\                        
39     & 64397.55$^{ }$   & 64276.49$^{ }$   & 64158.82$^{ }$   &                   &                   &                  \\                        
40     & 64375.72$^{ b }$ & 64251.90$^{ }$   & 64131.12$^{ }$   &                   &                   &                  \\                        
41     & 64353.41$^{ b }$ & 64226.49$^{ }$   &                  &                   &                   &                  \\                        
42     & 64330.26$^{ }$   & 64200.50$^{ }$   &                  &                   &                   &                  \\                        
43     & 64306.48$^{ w }$ & 64173.82$^{ }$   &                  &                   &                   &                  \\                        
44     & 64282.31$^{ w }$ & 64146.45$^{ }$   &                  &                   &                   &                  \\                        
45     & 64257.29$^{ w }$ & 64118.75$^{ }$   &                  &                   &                   &                  \\

      \hline
      \hline
    \end{tabular}
  \end{center}
\end{table*}

\vspace{2cm}

\begin{table*}
  \begin{center}
    \caption{Lines due to states perturbing \As$(0)$ identified in the VUV-FT spectra and VIS-FT spectra in \CO. Transition frequencies (in vacuum \wn). $J''$ is the rotational quantum number of the lower state and subscripts $e$ and $f$ indicate the perturber states parity. The superscripts \textit{b} and \textit{w} indicate blended and weak transitions, respectively. The left-superscript $o$, $p$, $q$, $r$ and $s$ denote the total angular momentum excluding spin of the perturber states, according to the notation in Ref.~\cite{Morton1994}.}
\label{FT-pert}
\footnotesize
\begin{tabular}
{c@{\hspace{3pt}}l@{\hspace{2pt}}l@{\hspace{2pt}}l@{\hspace{3pt}}l@{\hspace{2pt}}l@{\hspace{2pt}}l@{\hspace{3pt}}l@{\hspace{2pt}}l@{\hspace{2pt}}l}

\hline
\hline\\[-2ex]

$J''$        & \multicolumn{1}{c}{${}^q\!R_e$} & \multicolumn{1}{c}{${}^p\!Q_f$} & \multicolumn{1}{c}{${}^o\!P_e$} & \multicolumn{1}{c}{${}^r\!R_e$} & \multicolumn{1}{c}{${}^q\!Q_f$} & \multicolumn{1}{c}{${}^p\!P_e$} & \multicolumn{1}{c}{${}^s\!R_e$} & \multicolumn{1}{c}{${}^r\!Q_f$} & \multicolumn{1}{c}{${}^q\!P_e$}\\
\hline      \\[-2ex]
             &\multicolumn{9}{c}{\eX\ (1,0)}  \\
\cline{2-10}\\[-1.5ex]
0            & 64789.81$^{w}$                  &                                 &                                 &                                 &                                 &                                 & 64796.51                        &                                 &                                \\
1            & 64788.60                        &                                 &                                 &                                 & 64789.81$^{b}$                  &                                 & 64800.11                        &                                 & 64786.83$^{w}$                 \\
2            & 64786.23                        &                                 & 64778.84$^{wb}$                 &                                 & 64787.33                        &                                 & 64802.39                        &                                 & 64785.50$^{w}$                 \\
3            & 64782.70                        &                                 & 64770.23$^{wb}$                 &                                 & 64783.62                        &                                 & 64803.41                        &                                 & 64781.70$^{w}$                 \\
4            & 64778.11                        &                                 & 64760.49                        &                                 & 64778.70                        &                                 & 64803.16                        &                                 & 64776.70$^{w}$                 \\
5            & 64772.76                        &                                 & 64749.67$^{wb}$                 &                                 & 64772.59                        &                                 & 64801.61                        &                                 & 64770.30$^{w}$                 \\
6            & 64767.39                        &                                 & 64737.73                        &                                 & 64765.30                        &                                 & 64798.88                        &                                 & 64762.75$^{w}$                 \\
7            & 64749.29$^{b}$                  &                                 & 64724.99                        &                                 & 64756.96                        &                                 & 64794.85                        &                                 & 64753.85                       \\
8            & 64740.65                        &                                 & 64712.28                        &                                 & 64747.67$^{b}$                  &                                 & 64789.60                        &                                 &                                \\
9            & 64729.91                        &                                 & 64686.88                        &                                 & 64737.81                        &                                 & 64783.14                        &                                 &                                \\
10           &                                 &                                 & 64670.86                        &                                 & 64710.79                        &                                 & 64775.49                        &                                 &                                \\
11           & 64703.82$^{w}$                  &                                 & 64652.75                        &                                 & 64699.64                        &                                 & 64766.80                        &                                 & 64705.96                       \\
12           &                                 &                                 &                                 &                                 & 64686.15                        &                                 & 64757.25                        &                                 &                                \\
13           &                                 &                                 &                                 &                                 & 64670.70                        &                                 & 64747.67$^{b}$                  &                                 & 64674.96                       \\
14           &                                 &                                 &                                 &                                 & 64653.53                        &                                 & 64725.69                        &                                 & 64658.12                       \\
15           &                                 &                                 &                                 &                                 & 64634.85                        &                                 & 64712.88                        &                                 & 64641.16                       \\
16           &                                 &                                 &                                 &                                 & 64614.79                        &                                 & 64698.03                        &                                 & 64611.96$^{b}$                 \\
17           &                                 &                                 &                                 &                                 &                                 &                                 &                                 &                                 & 64591.78                       \\
\vspace{1ex}
             &\multicolumn{9}{c}{\dX\ (4,0)}  \\
\cline{2-10}\\[-1.5ex]
22           & 64674.24                        &                                 &                                 &                                 &                                 &                                 &                                 &                                 &                                \\
23           &                                 & 64589.95                        &                                 &                                 &                                 &                                 &                                 &                                 &                                \\
24           &                                 &                                 & 64502.05                        &                                 &                                 &                                 &                                 &                                 &                                \\
25           &                                 &                                 &                                 & 64654.61                        & 64592.36                        &                                 &                                 &                                 &                                \\
26           &                                 &                                 &                                 & 64612.78                        & 64559.42                        &                                &                                &                                &                                \\
27          &                                &                                &                                & 64584.16$^{w}$                 & 64514.02$^{b}$                 & 64460.57$^{w}$                 &                                &                                &                                \\
28          &                                &                                &                                &                                & 64481.73                       & 64411.54                       &                                &                                &                                \\
29          &                                &                                &                                &                                &                                &                                & 64601.35$^{w}$                 & 64527.31$^{w}$                 &                                \\
30          &                                &                                &                                &                                &                                &                                & 64554.18                       & 64491.67$^{w}$                 &                                \\
31          &                                &                                &                                &                                &                                &                                & 64522.49                       & 64440.87                       & 64378.36                       \\
32          &                                &                                &                                &                                &                                &                                &                                & 64405.60                       & 64324.06$^{b}$                 \\
33          &                                &                                &                                &                                &                                &                                &                                &                                & 64285.07                       \\
\vspace{1ex}
            &\multicolumn{9}{c}{\Be\ (0,1)}  \\
\cline{2-10}\\[-1.5ex]
4  &                   &                 &                    &                    & 22138.60$^{w}$  &                   &                    &                    &                  \\
5  &                   &                 &                    &                    & 22144.97$^{wb}$ &                   & 22180.14$^{wb}$ &                    & 22139.18$^{w}$\\
6  &                   &                 &                    &                    & 22152.53$^{w}$  &                   & 22193.17$^{b}$  &                    & 22144.76$^{b}$\\
7  &                   &                 &                    &                    & 22161.22$^{b}$  &                   & 22206.286$^{}$   &                    & 22150.438$^{}$ \\
8  &                   &                 &                    &                    & 22170.864$^{}$   &                   & 22232.13$^{wb}$ &                    & 22168.83$^{b}$\\
9  &                   &                 &                    &                    & 22181.188$^{}$   &                   &                    &                    & 22177.90$^{b}$\\
10 &                   &                 &                    &                    & 22208.695$^{}$   &                   &                    &                    &                  \\
11 &                   &                 &                    &                    & 22220.392$^{}$   &                   &                    &                    &                  \\
12 &                   &                 & 22153.25$^{w}$  &                    & 22234.46$^{b}$  &                   &                    &                    &                  \\
13 &                   &                 & 22163.36$^{b}$  &                    & 22250.62$^{wb}$ &                   &                    &                    &                  \\
14 & 22281.503$^{}$  &                 & 22173.63$^{b}$  &                    & 22268.41$^{wb}$ &                   &                    &                    &                  \\
15 & 22311.53$^{b}$ &                 & 22196.217$^{}$   &                    &                    &                   &                    &                    &                  \\
\vspace{1ex}
            &\multicolumn{9}{c}{\Bd\ (0,4)}  \\
\cline{2-10}\\[-1.5ex]
23 &                   &                 &                    &                    &                    &                   & 22429.30$^{b}$  & 22340.26$^{b}$  &                  \\
24 &                   &                 &                    &                    &                    &                   &                    & 22376.78$^{wb}$ &                  \\
25 &                   &                 &                    &                    & 22340.17$^{wb}$ &                   &                    &                    &                  \\
26 &                   &                 &                    &                    &                    &                   &                    &                    &                  \\
27 &                   &                 &                    & 22524.82$^{wb}$ & 22421.08$^{wb}$ & 22320.96$^{wb}$   &                    &                    &                  \\

\hline
\hline

\end{tabular}
%\end{scriptsize}
\end{center}
\end{table*}

\subsection{VIS-FT spectroscopy}

The \BA$(0,0)$ band was measured in the emission from a water-cooled hollow-cathode discharge lamp with two anodes that has been used and described in recent years
for a number of investigations of the visible spectra of the \AA ngstr\"{o}m~\cite{Hakalla2016} and Herzberg bands~\cite{Hakalla2014} for CO isotopologues.
The discharge lamp undergoes a preparation phase depositing either $^{12}$C or $^{13}$C atomic isotopes onto the electrodes after which the oxygen containing isotope
of choice is then introduced and a DC-plasma formed by applying a potential (650\,V) onto the anodes. In the discharge, specific isotopologues of CO are formed and maintained at a certain density and temperature (600$-$700\,K) for several hours, emitting visible radiation to be investigated by spectroscopic means.

The spectroscopic system, previously composed of an Ebert plane-grating spectrograph combined with a laterally scanned photo-multiplier tube, was upgraded for the present investigation with a Bruker-IFS 125 HR Fourier-transform spectrometer recording inteferograms in the double-sided acquisition mode. The CO discharge light source was focused on the entrance aperture of the FTS by a plano-convex quartz lens. The photomultiplier tube operated in integration pulse mode was used as an internal detector covering the 11\,000--25\,000 \wn\ spectral region. The instrumental resolution limit was set at 0.018 \wn. Higher resolution was not necessary, as the Doppler-dominated linewidth (FWHM) of the CO lines was about 0.06 \wn. A total of 126 scans were co-added and the spectrum has a good signal-to noise ratio of 120:1. The FTS was operated under vacuum conditions. 

The new VIS-FT setup allowed for a two-times improvement in the width of measured lines (0.08 \wn\,FWHM) as well as a better signal-to-noise ratio relative to the old grating spectrograph.
Previously our CO spectra were contaminated with a dense and intense coverage of Th frequency-calibration lines. This is no longer necessary given the intrinsically linear frequency scale of the Fourier-transform spectrometer and much cleaner spectra can be obtained.
A detailed description of the improved setup will be the subject of a future publication.

\begin{figure}
\begin{center}
\includegraphics[width=\textwidth]{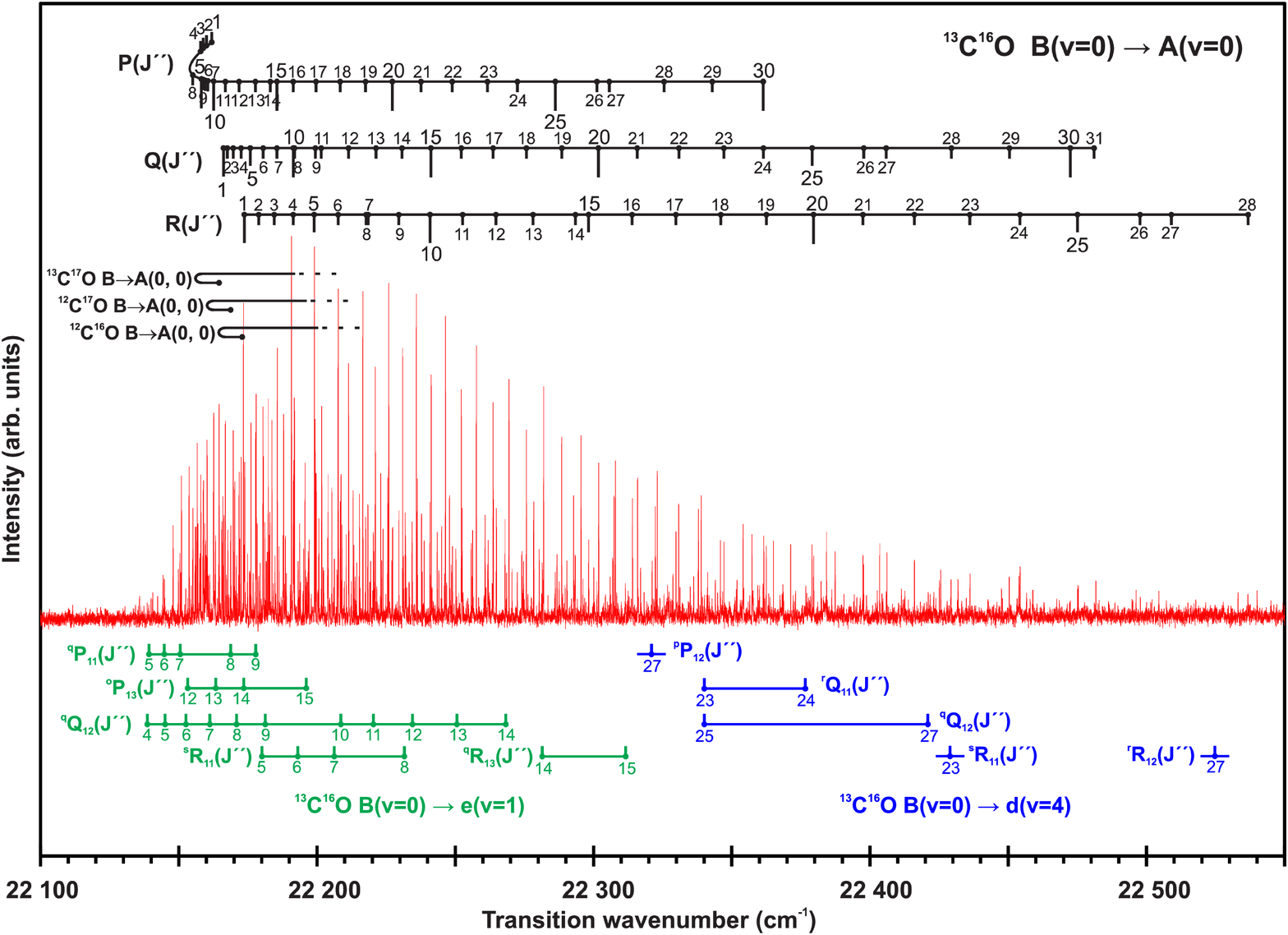}
\caption{High resolution emission spectra of the \CO\ \BA\ $(0,0)$, \Be\ $(0,1)$, and \Bd\ $(0,4)$ systems recorded with the VIS-FTS setup with an instrumental resolution of 0.018~\wn. The estimated accuracy of the frequency scale calibration is 0.003~\wn. Absolute accuracies of the frequency measurements vary between 0.003~\wn\ for the strongest lines and 0.03~\wn\ for the weakest or most blended. The composition ratio of the gas used in the experiment was $^{13}$C$^{16}$O : $^{13}$C$^{17}$O = 0.43 : 1.}
\label{fig:FTBXspec}
\end{center}
\end{figure}

In Fig.~\ref{fig:FTBXspec} a spectrum of the \BA$(0,0)$ band of $^{13}$C$^{16}$O is presented.
Also present are lines from the forbidden \Be$(1,0)$ and \Bd$(0,4)$ bands, which borrow intensity from \BA$(0,0)$ transition.
Line positions were measured by fitting Voigt lineshape functions to the experimental lines in a Levenberg-Marquardt procedure~\cite{Levenberg1944, Marquardt1963} included in the OPUS$^\textrm{TM}$  software package~\cite{OPUS}. 
An absolute calibration of the wavenumber scale accurate to within 0.003\,\wn was made with reference to the first harmonic of a frequency-stabilized He-Ne laser.
This laser monitors the FTS optical path difference and is stable over 1\,h of operation within a fractional variation of $\pm2\times10^{-9}$.

The frequencies of strong and unblended transitions were determined with a $1\sigma$ random uncertainty of 0.002\,\wn{}, and weak and blended lines with somewhat greater uncertainty. The absolute accuracy of the frequency measurements then varies between 0.003 and 0.03\,\wn{} for the strongest and weakest/blended lines, respectively.   

Transition frequencies for the \BA$(0,0)$ band of $^{13}$C$^{16}$O  are listed in Table~\ref{FTandlasertransition}, and frequencies of the perturber lines are listed in Table~\ref{FT-pert}.

\begin{table*}
\begin{center}
\caption{Term values of \As\,(0) and perturbing states. A {\large$\ast$} indicates level energies calculated from lines observed in the laser experiment.}
\label{Levels}
\renewcommand\arraystretch{1.1}
\footnotesize
\begin{tabular}
{c@{\hspace{25pt}}l@{\hspace{4pt}}l@{\hspace{25pt}}l@{\hspace{4pt}}l@{\hspace{8pt}}l@{\hspace{4pt}}l@{\hspace{8pt}}l@{\hspace{4pt}}l}
\hline
\hline\\[-2ex]
	&	\multicolumn{2}{c}{\As\,$(0)$}			&	\multicolumn{6}{c}{\es\ $(v=1)$}											\\
{$J$}&	\multicolumn{1}{c}{$e$}           &	\multicolumn{1}{c}{$f$}           &	\multicolumn{1}{c}{$F_1e$}	&	\multicolumn{1}{c}{$F_1f$}	&	\multicolumn{1}{c}{$F_2e$}	&	\multicolumn{1}{c}{$F_2f$}	&	\multicolumn{1}{c}{$F_3e$}	&	\multicolumn{1}{c}{$F_3f$}	\\
% \cline{2-9}\\[-2ex]
% \cline{1-3}
% \cline{4-9}\\[-2ex]
\hline\\[-2ex]

0  &                &               &		&		&		&		&	64790.51&		\\
1  &	64754.157\,*&	64754.170\,*&	64789.84&		&		&	64793.48&	64796.52&		\\
2  &	64760.184\,*&	64760.235\,*&	64792.28&		&		&	64798.36&	64803.77&		\\
3  &	64769.202\,*&	64769.32    &	64797.25&		&		&	64805.67&	64813.43&		\\
4  &	64781.16    &	64781.42    &	64804.77&		&		&	64815.45&	64825.45&		\\
5  &	64795.94    &	64796.50    &	64814.89&		&		&	64827.72&	64839.92&		\\
6  &	64813.28    &	64814.52    &	64827.90&		&		&	64842.49&	64856.75&		\\
7  &	64832.42    &	64835.42    &	64844.58&		&		&	64859.87&	64876.06&		\\
8  &	64865.97    &	64859.01    &	64852.24&		&		&	64879.98&	64897.76&		\\
9  &	64891.95    &	64885.02    &	64872.96&		&		&	64903.19&	64921.90&		\\
10 &	64921.69    &	64929.95    &	64895.27&		&		&	64912.90&	64948.49&		\\
11 &	64954.70    &	64960.80    &		&		&		&	64942.15&	64977.60&		\\
12 &	64990.71    &	64995.76    &	64946.33&		&		&	64972.74&	65009.30&		\\
13 &	65029.37    &	65034.48    &		&		&		&	65005.02&	65043.85&		\\
14 &	65069.98    &	65076.66    &		&		&		&	65039.26&	65081.97&		\\
15 &	65124.66    &	65122.11    &		&		&		&	65075.64&	65111.43&		\\
16 &	65172.00    &	65170.73    &		&		&		&	65114.31&	65153.68&		\\
17 &	65223.18    &	65222.46    &		&		&		&		&	65197.54&		\\
18 &	65277.73    &	65277.28    &		&		&		&		&		&		\\
19 &	65335.45    &	65335.15    &		&		&		&		&		&		\\
20 &	65396.28    &	65396.08    &		&		&		&		&		&		\\
21 &	65460.13    &	65459.96    &		&		&		&		&		&		\\
22 &	65526.92    &	65526.80    &	\multicolumn{6}{c}{\ds\ $(v=4)$}											\\
23 &	65595.86    &	65595.78    &	\multicolumn{1}{c}{$F_1e$}	&	\multicolumn{1}{c}{$F_1f$}	&	\multicolumn{1}{c}{$F_2e$}	&	\multicolumn{1}{c}{$F_2f$}	&	\multicolumn{1}{c}{$F_3e$}	&	\multicolumn{1}{c}{$F_3f$}	\\
\cline{4-9}\\[-2.5ex]               
24 &	65670.54    &	65670.46    &	65602.82&	65602.81&		&		&		&		\\               
25 &	65745.86    &	65745.80    &		&		&		&		&		&		\\               
26 &	65823.59    &	65823.54    &		&		&		&	65784.68&		&		\\               
27 &	65915.52    &	65915.49    &		&		&	65846.91&	65846.93&		&		\\               
28 &	65995.79    &	65995.75    &		&		&	65900.29&	65900.33&		&		\\               
29 &	66082.19    &	66082.14    &		&		&	65970.47&	65970.48&		&		\\               
30 &	66171.18    &	66171.15    &		&		&		&		&		&	66122.11	\\       
31 &	66277.23    &	66277.17    &		&		&		&		&	66196.13&	66196.15	\\       
32 &	66369.56    &	66369.51    &		&		&		&		&	66258.68&	66258.64	\\       
33 &	66468.29    &	66468.20    &		&		&		&		&	66340.25&	66340.27	\\       
34 &	66570.82    &	66570.90    &		&		&		&		&		&		\\               
35 &	66676.60    &	66676.62    &		&		&		&		&		&		\\               
36 &	66785.35    &	66785.50    &		&		&		&		&		&		\\               
37 &	66897.62    &	66897.35    &		&		&		&		&		&		\\               
38 &	67012.47    &	67012.32    &		&		&		&		&		&		\\               
39 &	67130.36    &	67130.14    &		&		&		&		&		&		\\               
40 &	67251.20    &	67251.16    &		&		&		&		&		&		\\               
41 &	67374.97    &	67374.92    &		&		&		&		&		&		\\               
42 &	67501.84    &	67501.66    &		&		&		&		&		&		\\               
43 &	67631.43    &	67631.28    &		&		&		&		&		&		\\               
44 &	67763.93    &	67763.75    &		&		&		&		&		&		\\               
45 &	67899.61    &	67899.43    &		&		&		&		&		&		\\               
46 &	68037.98    &               &		&		&		&		&		&		\\

\hline
\hline
\end{tabular}
\end{center}
\end{table*}

\section{Level energies}

The laser and VUV-FT transition frequencies were converted to \As\,(0), \es\,(1) and \ds\,(4) level energies using accurate level energies for the ground state~\cite{George1994}.
The combination of \As\,(0) level energies and \BA$(0,0)$ emission line frequencies also permitted the determination of \Bs\,(0) level energies. 
As a verification of our various frequency calibrations we compared the deduced \Bs\,$(0)$ energy levels to those calculated from an unpublished \BX$(0,0)$ photoabsorption spectrum also recorded at SOLEIL \cite{BX_unpublished}. 
Comparing level energies determined from B$\rightarrow$A and A$\leftarrow$X  spectra against this direct B$\leftarrow$X measurement we find agreement within 0.01\,\wn{}, which is the estimated uncertainty of our wavenumber calibration.

All \As\,(0), \es\,(1) and \ds\,(4) level energies are listed in Table~\ref{Levels} and represent weighted averages of the different experiments. 
The low-$J$ values are taken from the more accurate laser data.

The parallel ladders of $e$- and $f$-parity $J$-levels of \As$(0)$ are nondegenerate because of the influence of perturbing $\Sigma$ states.
The laser-measured $S$ and $Q$-branch lines provide information on $e$-parity \As\,$(0)$ levels, while the $P$ and $R$ branches provide information on the $f$-parity components.
Whereas, under the selection rules of one-photon absorption and emission, $R$ and $P$-branch lines probe $e$-parity levels, and the Q-branch lines probes $f$-parity components.
The \es\,(1) and \ds\,(4) levels are split into $F_1$, $F_2$, and $F_3$; and $e$- and $f$-parity levels, not all of which were accessed in our experiments.

\begin{table}
\caption{Compilation of molecular constants for the \As\,$(v=0)$ state of \CO\ and all perturber states from the deperturbation analysis. All values in vacuum \wn. Uncertainties ($1\sigma$) given in parentheses are in units of the least significant digit.}
\label{Molecons}

\begin{center}
\begin{tabular}{l@{\hspace{10pt}}r@{\hspace{10pt}}r@{\hspace{10pt}}r}

\hline
\hline\\[-1.5ex]

Constant              & A$^1\Pi (v=0)$        & B$^1\Sigma^+ (v=0)$    & D$^1\!\Delta (v=0)$   \\
\hline               \\[-2ex]
$T_v$                 &$ 64753.3690(8)       $ &$ 86916.808(2){}^{\rm a}         $ &$ 65434.7557{}^{\rm b}$\\
$B$                   &$ 1.533674(5)         $ &$ 1.862475(5){}^{\rm a}          $ &$ 1.19331{}^{\rm b}   $\\
$q$ $\times 10^{5}$   &$ 5.9(3)              $ &$                      $           &$                     $\\
$D$ $\times 10^{6}$   &$ 6.743(3)            $ &$ 6.114(2){}^{\rm a}             $ &$ 6.43{}^{\rm b}      $\\
$H$ $\times 10^{12}$  &$ -12{}^{\rm b}       $ &$ -9(3){}^{\rm a}         $        &$ -0.26{}^{\rm b}     $\\
$\xi$                 &$                     $ &$                      $           &$ -0.0256(9)          $\\
                     \\[-1.5ex]
Constant              & e$^3\Sigma^- (v=0)$   & d$^3\!\Delta (v=3)$    & a$'^3\Sigma^+ (v=8)$  \\
\hline               \\[-2ex]
$T_v$                 &$ 63716.208{}^{\rm b} $ &$ 63955.0056{}^{\rm b} $           &$ 64269.6767{}^{\rm b}$\\
$B$                   &$ 1.21916{}^{\rm b}   $ &$ 1.19670{}^{\rm b}    $           &$ 1.14804{}^{\rm b}   $\\
$A$                   &$                     $ &$ -16.05{}^{\rm b}     $           &$                     $\\
$\lambda$             &$ 0.485{}^{\rm b}     $ &$ 0.69{}^{\rm b}       $           &$ -1.136{}^{\rm b}    $\\
$\gamma$              &$                     $ &$ -0.008{}^{\rm b}     $           &$ -0.006{}^{\rm b}    $\\
$D$ $\times 10^{6}$   &$ 6.3{}^{\rm b}       $ &$ 5.94{}^{\rm b}       $           &$ 5.74{}^{\rm b}      $\\
$H$ $\times 10^{12}$  &$ -1.75{}^{\rm b}     $ &$ -0.70{}^{\rm b}      $           &$ -0.35{}^{\rm b}     $\\
$A_D$ $\times 10^{4}$ &$                     $ &$ -0.5{}^{\rm b}       $           &$                     $\\
$\eta$                &$ -9.410{}^{\rm c}    $ &$ 24.802{}^{\rm c}     $           &$ 3.60{}^{\rm c}      $\\
                     \\[-1.5ex]
Constant              & e$^3\Sigma^- (v=1)$   & d$^3\!\Delta (v=4)$    & a$'^3\Sigma^+ (v=9)$  \\
\hline               \\[-2ex]
$T_v$                 &$ 64788.758(1)        $ &$ 65017.4897(9)        $           &$ 65297.0683{}^{\rm b}$\\
$B$                   &$ 1.20223(4)          $ &$ 1.18059(4)           $           &$ 1.13274{}^{\rm b}   $\\
$A$                   &$                     $ &$ -16.43(1)            $           &$                     $\\
$\lambda$             &$ 0.525(4)            $ &$ 1.44(3)              $           &$ -1.129{}^{\rm b}    $\\
$\gamma$              &$                     $ &$ -0.009{}^{\rm b}     $           &$ -0.006{}^{\rm b}    $\\
$D$ $\times 10^{6}$   &$ 6.2(2)              $ &$ 5.75(5)              $           &$ 5.73{}^{\rm b}      $\\
$H$ $\times 10^{12}$  &$ -1.75{}^{\rm b}     $ &$ -0.70{}^{\rm b}      $           &$ -0.35{}^{\rm b}     $\\
$A_D$ $\times 10^{4}$ &$                     $ &$ -0.5{}^{\rm b}       $           &$                     $\\
$\eta$                &$ 14.783(3)           $ &$ -21.902(7)           $           &$ -2.54(4)            $\\
                     \\[-1.5ex]
Constant              & e$^3\Sigma^- (v=2)$   &                        &                       \\
\hline               \\[-2ex]
$T_v$                 &$ 65841.794{}^{\rm b} $ &$                      $           &$                     $\\
$B$                   &$ 1.18559{}^{\rm b}   $ &$                      $           &$                     $\\
$A$                   &$                     $ &$                      $           &$                     $\\
$\lambda$             &$ 0.519{}^{\rm b}     $ &$                      $           &$                     $\\
$\gamma$              &$                     $ &$                      $           &$                     $\\
$D$ $\times 10^{6}$   &$ 6.2{}^{\rm b}       $ &$                      $           &$                     $\\
$H$ $\times 10^{12}$ &$ -1.75{}^{\rm b}     $&$                      $          &$                     $\\
$A_D$ $\times 10^{4}$&$                     $&$                      $          &$                     $\\
$\eta$               &$ -17.376{}^{\rm c}   $&$                      $          &$                     $\\

\hline
\hline

\end{tabular}
  
\end{center}
\tabnote{${}^{\rm a}$ Incorporates data from Ref.~\cite{BX_unpublished}.}
\tabnote{${}^{\rm b}$ Constants for \es, \ds, and \as{} were obtained by isotopic scaling \cite{Lefloch-thesis,Hakalla2016} values taken from Ref.~\cite{Field-thesis}, apart from $H$ which is rescaled from Ref.~\cite{Lefloch1987}. All \Ds{} constants were rescaled from Ref.~\cite{kittrell1989}, apart from $H$ from Ref.~\cite{Lefloch1987}. Equilibrium energies for the \ds{} and \es{} states were taken from Refs.~\cite{Herzberg1970} and \cite{Tilford1972}, respectively.}
\tabnote{${}^{\rm c}$ Estimated on the basis of the isotopologue-independent purely electronic perturbation parameters of Refs.~\cite{Field1972a, Field1972,Lefloch1987}.}
\end{table}

\section{Perturbation analysis}
The early analyses of Field \emph{et al.} \cite{Field1972} showed that the \As\ state of CO is perturbed by a multitude of vibrational levels pertaining to different electronic states, including \ds, \es, \as, \Ds, and \Is. Higher resolution and signal-to-noise spectra provide the opportunity to improve upon the original perturbation analyses, as was shown in recent studies for the main $^{12}$C$^{16}$O isotopologue~\cite{Niu2013,Niu2016} and for the $^{12}$C$^{17}$O isotopologue~\cite{Hakalla2016}. Specifically, for the \As\,$(0)$ level of $^{13}$C$^{16}$O, a perturbation analysis was previously performed from observations of the \AX\ system by Haridass and Huber \cite{Haridass1994} and by Gavilan \emph{et al.} \cite{Gavilan2013}, while K{\c e}pa \emph{et al.} \cite{Kepa2003} investigated the perturbation in \As\,$(0)$ by analyzing the \BA\ emission system. In these studies, the following perturber states for \As\,(0) state were identified: \es\,(1), \ds\,(4), and \as\,(9). The \ds\,(4) state of \CO\ was also detected in a separate experiment performed at higher column density \cite{Herzberg1970}.

\begin{figure}
\begin{center}
\includegraphics[width=\linewidth]{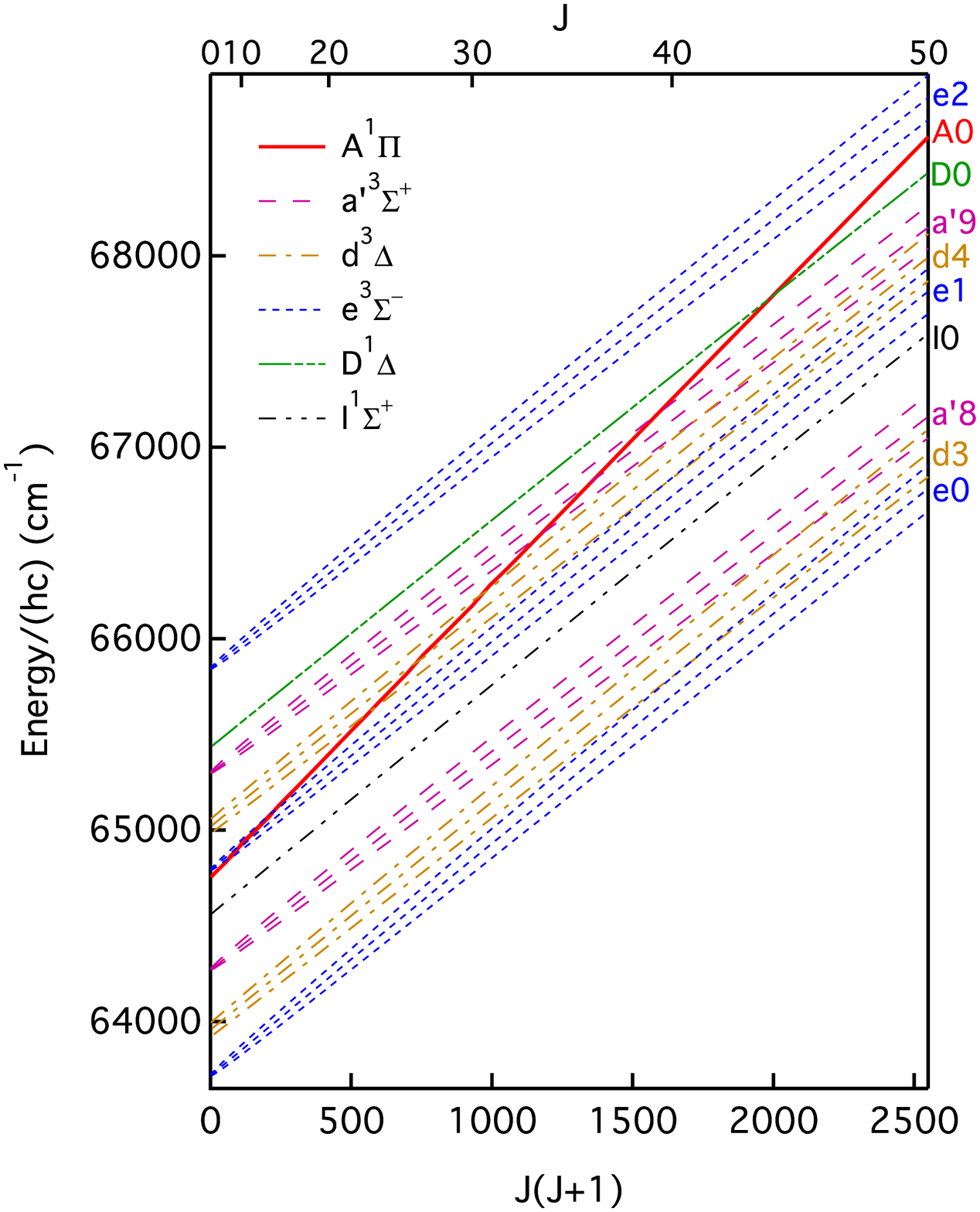}
\caption{Level energies as a function of $J(J+1)$ for \As\,(0) and its perturbers. The curves are indicated by an electronic state label and vibrational quantum number.}
\label{fig:Pert-diagram}
\end{center}
\end{figure}

Figure~\ref{fig:Pert-diagram} shows the progression of rotational level energies as a function of $J(J+1)$ for \As\,(0) and its possible perturber states in \CO, with data collected from our results and the previous works referenced above. This figure shows crossing points of \As\,(0) with  \es\,(1), \ds\,(4), ${\rm a}'\,{}^3\Sigma^+(9)$, and \Ds\,(0)   where localised perturbations are expected to occur. The perturber states \es\,$(v=0,2)$, \ds\,$(3)$, \as\,$(8)$ and \Is\,$(0)$ which do not cross the \As\,(0) state may still have some effects, which will be discussed below.

A similar methodology as in previous works \cite{Niu2013, Niu2016} is used in the present analysis of the perturbations affecting the \As\,$(0)$ state of $^{13}$C$^{16}$O.
Using the Pgopher software~\cite{pgopher}, we simulate all measured transition frequencies with the
same effective Hamiltonian model previously adopted in Ref.~\cite{Niu2013}. 
The molecular constants in this model were initialised by isotope-scaling the values deduced by Le~Floch \cite{Lefloch-thesis} for $^{12}$C$^{16}$O, and then optimised to best fit the experimental data.
A final fit was obtained after observing the correlated effect of pairs of molecular constants, and relied on a total of 17 independent molecular constants fitted to 397 transition frequencies.

The  transition frequencies obtained from VUV-FT spectra for levels up to $J'=46$ of the \AX\ $(0,0)$ band and perturber states are used in this analysis, as well as the frequencies from VIS-FT spectra of the \BA\ $(0,0)$ band. The greatest-accuracy transition frequencies from the laser experiment for low$-J$ of \AX\ $(0,0)$ transitions are preferentially used where possible. To perform a comprehensive perturbation analysis, we also use \dX\ $(4,0)$ frequencies from previous work \cite{Herzberg1970}.
In addition, to integrate the various data into our analysis, we weighted their experiment-model residual difference according to the inverse of their respective uncertainties. 
The unweighted root-mean-square of residuals from the final fit is 0.06 \wn, and is dominated by uncertainties of the lines taken from the least-accurate source, Ref~\cite{Herzberg1970}. The root-mean-square of weighted residuals was 1.2 and sufficiently close to 1 to indicate a satisfactory fit of the model and experiment as a whole.

The final set of deperturbed molecular constants obtained from the fit are listed in Table~\ref{Molecons}, where their definitions are the same as in Ref.~\cite{Niu2013}. The spin-orbit interaction parameters between \As\,(0) and triplet perturber levels are denoted by $\eta$, and the L-uncoupling interaction between \As\,(0) state and singlet perturbers by $\xi$. Constants with listed uncertainties were fitted parameters while all others were taken from sources indicated in the table footnotes (usually equilibrium constants found by studying other CO isotopologues) and fixed during the fitting procedure.
We included \es\,$(v=0,2)$, \ds\,(3) and \as\,(8) among our deperturbed levels, even though they have no crossings with \As\,$(0)$ within our observed $J$-range.
Their perturbative effects are expected to be significant but could not be independently estimated because of strong correlations with the \As\,$(0)$ band origin parameter, $T_v$.
Their interaction energies were then deduced from isotope-independent perturbation parameters originally based on experimental work in other isotopologues \cite{Lefloch1987,Hakalla2016} and adapted to interacting levels in our $^{13}$C$^{16}$O model by calculating the relevant isotope-dependent vibrational-overlap factors, as described in Hakalla \emph{et al.} \cite{Hakalla2016}.

Perturbation with the \Is$(0)$ state was also investigated in the analysis.
However, no effect on the model-experiment residuals could be determined for any reasonable L-uncoupling interaction between \Is\ $(0)$ and \As\,$(0)$, despite its strong $J$-dependence.
Because our spectra only provide data up to $J=46$, the effect of the \As\,$(0)$ sextic centrifugal distortion parameter, $H$, is at or below our measurement sensitivity and was fixed to the isotopically recalculated value from Le~Floch \cite{Lefloch1987} in our final fit.

\section{Conclusion}

New spectroscopic data on the \As\,$(0)$ level in $^{13}$C$^{16}$O was obtained by three different experimental techniques: \AX(0,0) 2+1 UV photon laser REMPI, \AX$(0,0)$ synchrotron absorption  and VUV Fourier-transform spectroscopy, and \BA$(0,0)$ photoemission in a discharge with visible  Fourier-transform spectroscopy.
The uncertainty of measured transition frequencies was 0.002\,\wn{} for the laser data, 0.01--0.03\,\wn{} for the VUV-FTS experiment, and 0.003--0.03 for the VIS-FTS experiment.
From this combined data, the level energies of \As\,(0) were determined up to $J=46$, as well as for the perturbing levels \es\,$(v)$ and \ds\,$(v)$, with accuracies between 0.002 to 0.03\,\wn.

Combining the new information with literature data on $^{13}$C$^{16}$O and other CO isotopologues allowed for a deperturbation of this complex of interacting levels and the fitting of model energy levels to all experimental data within their uncertainties.
This required fitting 17 independent adjustable parameters.
These newly-deduced parameters comprise the molecular constants governing the \As\,(0), \es(1), and \ds\,(4) levels, as well as the spin-orbit and L-uncoupling parameters connecting \es(1), \ds\,(4), \Ds\,(0) and ${\rm a}'\,{}^3\Sigma^+(9)$ with \As\,(0).

This data is part of an ongoing study of \As{} levels and their perturbations in multiple CO isotopologues.
It is our goal to establish isotope-independent (electronic) perturbation parameters to describe the \As{} of CO for all isotopologues, based on values derived in the present study for $^{13}$C$^{16}$O and previous works on  $^{12}$C$^{16}$O ~\cite{Niu2013,Niu2016}  and $^{12}$C$^{17}$O \cite{Hakalla2016}.

\section*{Acknowledgements}
R. Hakalla thanks Laserlab-Europe for support of this research (grants no. 654148 within the European Union's Horizon 2020 Research and Innovation Programme as well as no. 284464 within the EC's Seventh Framework Programme).
We are grateful to the general and technical staff of SOLEIL for providing beam time under project no. 20120653 and 20110191.
We thank J.-L. Lemaire, M. Eidelsberg, S. Federman, G. Stark, J.R. Lyons for sharing data on CO 
B$-$X$(0,0)$ photoabsorption ahead of its publication \cite{BX_unpublished}.

% \bibliographystyle{apsrev4-1}
%\bibliographystyle{tMPH}
%\bibliography{CO}

\end{document}